\documentclass[%
 reprint,
 amsmath,amssymb,
 aps,
]{revtex4-2}

\usepackage{graphicx}
\usepackage{dcolumn}
\usepackage{bm}

\usepackage{color}
\usepackage{subfigure}
\usepackage{bm}

\newcommand{\diag}{{\tt diag}}
\newcommand{\real}{{\tt real}}
\newcommand{\adam}{{\tt Adam}}
\newcommand{\denoiser}{{\tt denoiser}}

\begin{document}


\title{Quantitative phase retrieval for Zernike phase-contrast microscopy}

\author{Rikimaru Kurata$^1$}
\altaffiliation{The authors contributed equally to this work.}

\author{Keiichiro Toda$^{2,3}$}
\altaffiliation{The authors contributed equally to this work.}

\author{Genki Ishigane$^2$}

\author{Makoto Naruse$^1$}

\author{Ryoichi Horisaki$^1$}
\email{horisaki@g.ecc.u-tokyo.ac.jp}

\author{Takuro Ideguchi$^{2,3}$}
\email{ideguchi@ipst.s.u-tokyo.ac.jp}

\affiliation{$^1$Department of Information Physics and Computing, Graduate School of Information Science and Technology, The University of Tokyo, 7-3-1 Hongo, Bunkyo-ku, Tokyo 113-8656, Japan}%

\affiliation{$^2$Department of Physics, Graduate School of Science, The University of Tokyo, 7-3-1 Hongo, Bunkyo-ku, Tokyo 113-0033, Japan}%

\affiliation{$^3$Institute for Photon Science and Technology, The University of Tokyo, 7-3-1 Hongo, Bunkyo-ku, Tokyo 113-0033, Japan}%

\date{\today}

\begin{abstract}
We present a single-image numerical phase retrieval method for Zernike phase-contrast microscopy~(ZPM) that addresses halo and shade-off artifacts, as well as the weak phase condition, without requiring hardware modifications. By employing a rigorous physical model of ZPM and a gradient descent algorithm for its inversion, we achieve quantitative ZPM imaging. Our approach is experimentally validated using biological cells and its quantitative nature is confirmed through comparisons with digital holography observations.
\end{abstract}

\maketitle

\begin{figure}[t]
\begin{center}
	\includegraphics[scale=1.3]{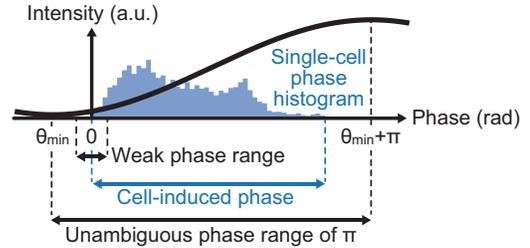}
\end{center}
\caption{An unambiguous phase range of $\pi$ and a weak phase range around 0 rad in ZPM.
The solid sinusoidal curve represents a contrast map for negative phase-contrast.
The blue bins show a histogram of a typical single-cell phase image.}
\label{fig_1}
\end{figure}

Phase-contrast microscopy, invented by Frits Zernike, has been widely used in biomedical fields for decades due to its simple implementation, which consists of a standard bright-field microscope with a condenser annulus and a phase ring attached~\cite{bib_zernike1955Phase}.
Although it provides high-contrast images of unstained transparent specimens, such as cells and tissues, Zernike phase-contrast microscopy~(ZPM) inevitably suffers from the halo and shade-off artifacts caused by the finite radial width of the phase ring, leading to its intrinsically non-quantitative nature~\cite{bib_Johnson2013Phase}.
Furthermore, although it is recognized that ZPM is nearly quantitative in the weak phase range for a condenser annulus and a phase ring with infinitely narrow radial widths~\cite{bib_goodman1996FourierOptics}, most biological specimens fall outside of this range because a single cell already induces a phase delay of a few rad, as shown in Fig.~\ref{fig_1}.
Therefore, ZPM is not well-suited for various biomedical analyses enabled by quantitative phase images, such as cellular dry mass or growth rate analyses, which have been intensively explored in recent studies on quantitative phase imaging~(QPI), including interferometric techniques like digital holography~(DH)~\cite{bib_Park2018Phase, bib_Nguyen2022Phase}. 

To address the aforementioned issues, a numerical approach for suppressing the artifacts of ZPM has been demonstrated~\cite{bib_Yin2012Phase}.
However, this method only considers the weak phase range, and the capability of quantitative phase retrieval for larger phases remains unexplored.
Alternative approaches have involved hardware modifications to the ZPM setup, enabling quantitative phase retrieval beyond the weak phase range.
For instance, spatial light interference microscopy (SLIM) is one of the representative techniques that realizes QPI~\cite{bib_Nguyen2017Halo}.
However, this technique requires an additional optical unit containing a spatial light modulator and relies on multiple phase-shifted measurements for the purpose of phase retrieval, which compromises the simplicity and robustness of ZPM.

In this study, we present a numerical method for quantitative phase retrieval from a single ZPM image without requiring hardware modification.
We revisit a physical model of ZPM and develop an efficient phase retrieval algorithm that robustly works for single-adherent-cell images.
As our technique does not necessitate any hardware modifications, it can be applied to any commercially available ZPM systems, thereby enabling effortless QPI.
This method may also be useful for other phase-contrast imaging modalities, such as X-ray or electron beam imaging~\cite{bib_Holzner2010Phase, bib_Danev2001Electron}.

In order to determine the available range for phase retrieval without ambiguity, it is essential to reconsider the physical model of ZPM.
The intensity $I^{\pm}$ of images acquired with positive and negative-contrast ZPMs, comprising a condenser annulus and a phase ring with infinitely narrow radial widths, can be written as a function of the object-induced phase delay $\theta$, given by
\begin{equation}
\begin{split}
I^{\pm}&=|\mp j\alpha+(\exp(j\theta)-1)|^2 \\
    &\propto I_0 + \sin(\theta-\pi/2-\theta_\text{min}^{\pm}),
\label{eq_ZPM}
\end{split}
\end{equation}
where $j$ is the imaginary unit, and $\alpha^2$ is the transmittance of the phase ring.
$I_0=(2+\alpha^2)/2\sqrt{1+\alpha^2}$ and $\theta_\text{min}^{\pm}$ that satisfies $\sin{\theta_\text{min}^{\pm}}=\pm\alpha/\sqrt{1+\alpha^2}$ and $\cos{\theta_\text{min}^{\pm}}=1/\sqrt{1+\alpha^2}$ are constant values characterized by the optical system, as depicted by the curve in Fig.~\ref{fig_1}.
This equation simply expresses interference between diffracted and non-diffracted light with a phase delay induced by the phase ring, resulting in a sinusoidal map with an unambiguous phase range of $\pi$.
It implies that single-image phase retrieval is feasible beyond the weak phase condition when the object-induced phase delays lie within the unambiguous phase range.
The minimum phase $\theta_\text{min}^{\pm}$ takes a value within the range of $0\leq\theta_\text{min}^{+}\leq\pi/4$ and $-\pi/4\leq\theta_\text{min}^{-}\leq0$ for positive and negative phase-contrast, respectively.
For weak phase objects that satisfy $\theta \ll$ 1 rad, Eq.~(\ref{eq_ZPM}) can be approximated as
\begin{equation}
\begin{split}
I^{\pm}&\simeq|\mp j\alpha+j\theta|^2 \\
    &\propto \alpha (\alpha\mp2\theta),
\label{eq_ZPM_weak}
\end{split}
\end{equation}
where the contrast of ZPM exhibits linearity with respect to $\theta$. However, it should be noted that most samples fail to meet this requirement, and the non-linear relationship in Eq.~(\ref{eq_ZPM}) must be considered.

Although we showed the potential for overcoming the weak phase range in ZPM without hardware modification, the issue of the halo and shade-off artifacts still remains because the condenser annulus and the phase ring are not infinitely narrow in practice.
To solve this issue, a numerical propagation model that rigorously describes the spatially partially-coherent light from the condenser annulus is needed, which would impose a tremendous computational cost.
We address this problem using a stochastic gradient descent approach called compressive propagation~(CP) proposed in our recent work~\cite{bib_Horisaki2022Coherence}.

\begin{figure}[t]
\begin{center}
	\includegraphics[scale=0.7]{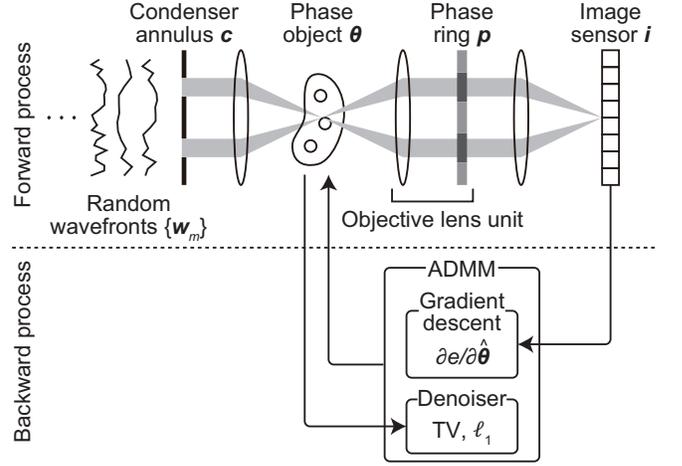}
\end{center}
\caption{Schematic diagram of the forward and backward processes for phase retrieval in ZPM.}
\label{fig_flow}
\end{figure}

For simplicity, we assume a one-dimensional phase object vector~$\bm{f}\in\mathbb{C}^{N\times 1}$, where $N$ is the pixel count of the object, as
\begin{equation}
\bm{f}=\exp(j\bm{\theta}),
\label{eq_obj}
\end{equation}
where $\bm{\theta}\in\mathbb{R}^{N\times 1}$ is the phase vector.
As shown in Fig.~\ref{fig_flow}, the forward model of ZPM based on CP is written as
\begin{equation}
\bm{i}=\frac{1}{M}\sum_{m=1}^M|\bm{F}^{-1}\diag(\bm{p})\bm{F}\diag(\bm{f})\bm{F}^{-1}\diag(\bm{c})\bm{w}_m|^2,
\label{eq_forward}
\end{equation}
where $\bm{i}\in\mathbb{R}^{N\times 1}$ is the vector of the captured intensity image, $\bm{p}\in\mathbb{C}^{N\times 1}$ is the vector of the phase ring, and $\bm{c}\in\mathbb{R}^{N\times 1}$ is the vector of the condenser annulus, respectively.
$\bm{w}_m\in\mathbb{C}^{N\times 1}$ is the vector of the $m$-th wavefront in $M$~random wavefronts.
$\bm{F}\in\mathbb{C}^{N\times N}$ and $\bm{F}^{-1}\in\mathbb{C}^{N\times N}$ denote matrices for the Fourier transform and its inversion, respectively.
``$\diag$'' is an operator for generating a diagonal matrix, where the diagonal elements are the parenthesized vector. 
As shown in the forward model in Eq.~(\ref{eq_forward}),  spatially partially-coherent light from the condenser annulus~$\bm{c}$ is expressed by an ensemble average of random wavefronts~$\bm{w}_m$ based on CP for reducing the computational cost in the gradient descent process mentioned bellow.

To perform gradient descent for phase retrieval in ZPM as shown in Fig.~\ref{fig_flow}, we define the error function~$e$ as
\begin{equation}
e=\|\widehat{\bm{i}}-\bm{i}\|_2^2,
\label{eq_error}
\end{equation}
where the accent mark of $\widehat{\bullet}$ denotes estimated variables during the gradient descent process and $\|\bullet\|_2$ is the $\ell_2$~norm, respectively.
The partial derivative of the error~$e$ in Eq.~(\ref{eq_error}) with respect to the estimated phase vector~$\widehat{\bm{\theta}}$ based on the chain rule is written as
\begin{equation}
\frac{\partial e}{\partial \widehat{\bm{\theta}}}=\frac{\partial \widehat{\bm{f}}}{\partial \widehat{\bm{\theta}}}\cdot\frac{\partial e}{\partial \widehat{\bm{f}}}.
\label{eq_grad}
\end{equation}
The second term of the right side in Eq.~(\ref{eq_grad}) is calculated as
\begin{equation}
\begin{split}
\frac{\partial e}{\partial \widehat{\bm{f}}}=&\frac{4}{M}\sum_{m=1}^M\diag(F^{-1}\diag(\bm{c})\bm{w}_m)^H\bm{F}^{-1}\diag(\bm{p})^H\bm{F}\\
&\diag(\bm{F}^{-1}\diag(\bm{p})\bm{F}\diag(\widehat{\bm{f}})\bm{F}^{-1}\diag(\bm{c})w_m)(\widehat{\bm{i}}-\bm{i}),
\end{split}
\label{eq_grad_f}
\end{equation}
where the superscript~$H$ denotes the Hermitian conjugate of a matrix.
Then, the left side of Eq.~(\ref{eq_grad}) is calculated as
\begin{equation}
\frac{\partial e}{\partial \widehat{\bm{\theta}}}=\real\left[-j\diag(\widehat{\bm{f}})^H\frac{\partial e}{\partial \widehat{\bm{f}}}\right],
\label{eq_grad_theta}
\end{equation}
where ``$\real$'' denotes the real part of the complex amplitudes.
We update the estimated phase vector~$\widehat{\bm{\theta}}^{(k)}$ at the $k$-th iteration with gradient descent based on the Adam optimizer as
\begin{equation}
\widehat{\bm{\theta}}^{(k+1)}=\widehat{\bm{\theta}}^{(k)}-\adam\left[\frac{\partial e}{\partial \widehat{\bm{\theta}}^{(k)}}\right],
\label{eq_adam}
\end{equation}
where ``$\adam$'' is an operator of the Adam optimizer for calculating the updating step with the partial derivative in Eq.~(\ref{eq_grad_theta})~\cite{bib_Kingma2015Adam}.
In CP, we randomly change the wavefronts~$\bm{w}_m$ in Eq.~(\ref{eq_grad_f}) at each iteration based on stochastic gradient descent~\cite{bib_Horisaki2022Coherence}.
The number of random wavefronts, $M$, can be kept small in order to reduce the computational cost associated with the spatially-partially light propagation, without introducing any approximations.

Additionally, we introduced a regularization process based on the alternating direction method of multipliers~(ADMM), as shown in Fig.~\ref{fig_flow}~\cite{bib_Chan2017ADMM}.
In this case, the updating step in Eq.~(\ref{eq_adam}) is rewritten with two auxiliary vectors~$\bm{v}^{(k)}\in\mathbb{R}^{N\times 1}$ and $\bm{u}^{(k)}\in\mathbb{R}^{N\times 1}$ at the $k$-th iteration as
\begin{align}
\widehat{\bm{\theta}}^{(k+1)}&=\widehat{\bm{\theta}}^{(k)}-\adam\left[\frac{\partial e}{\partial \widehat{\bm{\theta}}^{(k)}}+\rho(\widehat{\bm{\theta}}^{(k)}-(\bm{v}^{(k)}-\bm{u}^{(k)}))\right],
\label{eq_admm_x}\\
\bm{v}^{(k+1)}&=\denoiser\left[\widehat{\bm{\theta}}^{(k+1)}+\bm{u}^{(k)}\right],
\label{eq_admm_v}\\
\bm{u}^{(k+1)}&=\bm{u}^{(k)}+(\widehat{\bm{\theta}}^{(k+1)}-\bm{v}^{(k+1)}),
\label{eq_admm_u}
\end{align}
where $\rho$ is a tuning parameter.
Here, ``$\denoiser$'' is a denoising operator, which is composed of the total variation~(TV) to guarantee the smoothness of the object while preserving edges, and the $\ell_1$~norm to suppress background noise in this study~\cite{bib_rudin1992TV}.
Both the TV and $\ell_1$~norm were implemented with the reweighting method to adaptively enhance the sparsity on the regularization domains~\cite{bib_candes2007Reweighted}.

To experimentally demonstrate our phase retrieval method, a ZPM system was implemented for single-cell imaging using a commercial microscope~(Olympus IX73) equipped with a 525-nm LED~(Thorlabs SOLIS-525C), a condenser annulus~(Ph2), a negative phase-contrast objective~(UPlanFLN 40x/0.75NHPh2) and a CMOS image sensor~(Basler acA2440-75um).
To build a numerical physical model that accurately describes our implemented ZPM system, it is essential to carefully set the model parameters, such as the radii and the radial widths of the condenser annulus and phase ring, as well a the transmittance of the latter. 
In our case, we experimentally measured these parameters using the following procedure (see~\cite{bib_Nguyen2017Halo} for more details). We imaged the back aperture of the objective onto the camera plane both with and without condenser annulus. In the former case, the size of the condenser annulus was determined, while in the latter case, the radius and transmittance of the phase ring $\alpha$ = 0.26 could be characterized from the attenuated area of the phase ring. The inner and outer radii of the condenser annulus and the phase ring were calculated as ratios to the numerical aperture of the objective lens, specifically as 0.31 and 0.40, and 0.33 and 0.38, respectively.

In the phase retrieval process, the number of random wavefronts, $M$, was set to 10.
Cells' silhouettes were segmented from ZPM images by detecting and dilating edges~\cite{bib_Gonzalez2018Morphology} and were utilized as the initial guesses for the phase retrieval.
In the Adam optimizer, we set the learning rate to 0.016, and the other parameters were the same as those in the original work~\cite{bib_Kingma2015Adam}.
In the ADMM, the tuning parameter~$\rho$ was set to 20.

\begin{figure}[t]
\begin{center}
	  \includegraphics[scale=0.55]{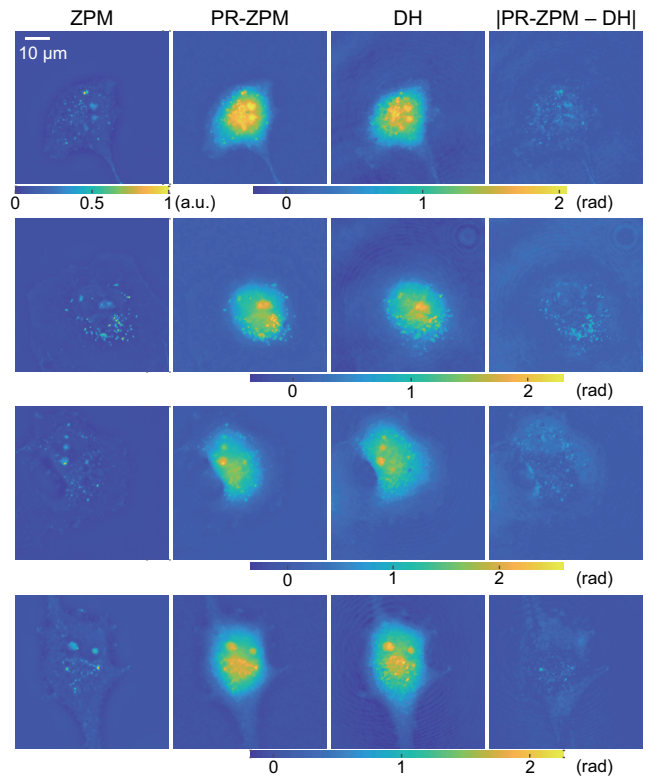}
\end{center}
\caption{Phase retrieval of COS-7 cell images captured with ZPM.
The first and second left columns show ZPM and their phase-retrieved (PR-ZPM) images, respectively.
The third and fourth columns show quantitative phase images independently captured with DH and the absolute difference between PR-ZPM and DH images, respectively. 
Phase values of DH images are determined relative to their background images.}
\label{fig_2}
\end{figure}

The first and second left columns in Fig.~\ref{fig_2} represent ZPM images of fixed COS-7 cells and their phase-retrieved ZPM (PR-ZPM) images, where the former are captured with our ZPM system. 
The half-pitch spatial resolution of ZPM images is 260 nm.
Our choice of negative phase-contrast guarantees full coverage of the range of interest~(0--a few rad) with the unambiguous phase range, as shown in Fig.~\ref{fig_1}.
As depicted in the figure, the phase retrieval results exhibit minimal residual halo and shade-off artifacts.
It is worth noting that the model and algorithmic parameters were fixed for the phase retrieval of all the cell images.

We compared the PR-ZPM images with quantitative phase images independently captured with a homemade DH system based on the common-path broadband diffraction phase microscopy technique \cite{bib_bhaduri2012Phase}, equipped with a 532-nm laser for illumination, an objective (LUCPLFLN40X, 40x/0.6), and the same image sensor as employed in the ZPM system. 
The half-pitch spatial resolution of DH images is 440 nm.
Details on our DH system are described in our recent work \cite{bib_Ishigane2022Photothermal}.  
As depicted in Fig.~\ref{fig_2}, the phase values of all DH images lie between 0 and approximately 2 rad, which largely surpasses the weak phase range.
The rightmost column in Fig.~\ref{fig_2} shows the absolute difference between PR-ZPM and DH images.
The standard deviations of the difference images within the intracellular regions, which are determined by the cells' silhouettes used for the initial guesses in the phase retrieval, are 0.18, 0.25, 0.25, and 0.16 rad, respectively, from top to bottom.
These small residuals verify the quantitative nature of our single-image phase retrieval algorithm based on an approximation-free model.

\begin{figure}[t]
\begin{center}
        \includegraphics[scale=0.55]{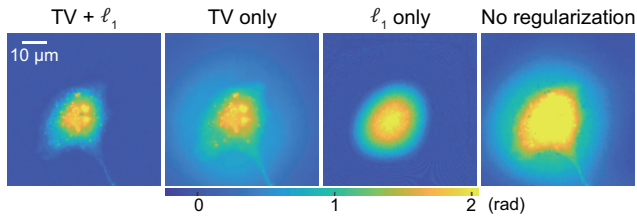}
\end{center}
\caption{Comparison of PR-ZPM images with different conditions in regularization.
The first left image is the result with both the TV and the $\ell_1$~norm, which is shown in the first row of Fig.~\ref{fig_2}.
The second and third images are the results with the TV only and with the $\ell_1$~norm only, respectively.
The fourth image is the result without the TV or the $\ell_1$~norm.}
\label{fig_4}
\end{figure}

To assess the effectiveness of regularization, we compared the phase-retrieved results obtained using different regularization methods.
Figure~\ref{fig_4} shows results with the TV and the $\ell_1$~norm, with the TV only, with the $\ell_1$~norm only, and without the TV or the $\ell_1$~norm, respectively.
As shown in these results, the TV contributed to reconstruct morphological structures of the cells, and the $\ell_1$~norm suppressed background noise.
The TV and the $\ell_1$~norm cooperatively functioned in the phase retrieval process.
The standard deviations of the absolute difference within the intracellular regions compared to the DH image in Fig.~\ref{fig_2} are 0.18, 0.47, 0.43, and 1.11 rad, respectively.

Finally, we discuss the possible cause of the residual phase discrepancies between PR-ZPM and DH images shown in Fig.~\ref{fig_2}. A mismatch in spatial resolution between the two systems may partially contribute to the residual, even though we adjusted the pixel pitch of the PR-ZPM images to match that of the DH images. Another factor could be the spatial phase noise of the DH measurements, primarily arising from the coherent noise due to laser illumination. We evaluated the standard deviations of the spatial phase noise in an area where no sample exists in the DH images and found values ranging from 0.02 to 0.04 rad, which are lower than the observed residuals. Consequently, we attribute the dominant cause of the residual to model error, which could be reduced through more precise determination of the system parameters.

In summary, we have developed a numerical phase retrieval method that allows quantitative ZPM in an unambiguous phase range of $\pi$ and successfully applied this method to single-cell imaging. 
By further reducing model errors, more accurate phase retrieval is expected.
Since our method can be implemented to any existing ZPM systems without hardware modification, it could open the door to the widespread use of QPI.

This work was supported by Japan Society for the Promotion of Science~(JP20H02657, JP20K05361, JP20H05890, JP20H00125, JP23H01874, JP23H00273), Asahi Glass Foundation, Research Foundation for Opto-Science and Technology, Nakatani Foundation, UTEC-UTokyo FSI Research Grant, and UTokyo IXT Project Support Program.


\bibliography{apssamp}

\end{document}